# THE DYNAMIC OF CONTRAST AGENT AND SURROUNDING FLUID IN THE VICINITY OF A WALL FOR SONOPORATION

**Nima Mobadersany, Kausik Sarkar**

**Department of Mechanical and Aerospace Engineering, The George Washington University,**

**Washington DC, USA**

**Abstract**
Contrast microbubbles (contrast agents) were initially developed to increase the contrast of the image in ultrasound imaging. These microbubbles consist of a gas core encapsulated by a layer of protein or lipid to stabilize them against early dissolution in the bloodstream. Contrast agents, in the presence of ultrasound, can also help facilitate the uptake of drugs and genes in to desired cells through the process called sonoporation. Sonoporation is the temporarily rupture of cell membranes in the presence of ultrasound. In this work, we have studied the contrast agent near a rigid wall (assumed as a cell membrane with high elastic modulus) in the presence of ultrasound using boundary element method. The contrast agent forms non-symmetrical high velocity microjet at the last stage of the collapse phase. The microjet and the adjacent surrounding fluid move toward the rigid wall with a very high velocity. This high velocity fluid impinges the wall and spreads radially along it. This will generate high velocity gradient on the wall which gives rise to shear stress resulting in the perforation of cell membrane. The encapsulation of the microbubble can be assumed as an interface with an infinitesimal thickness. There are several models to simulate the interface. In this study, the encapsulation is simulated with two models; viscoelastic model with exponentially varying elasticity (EEM) and Marmottant model. In this research, we have studied the effect of different parameters on the dynamics of the contrast microbubble, the induced shear stress on the wall due to its collapse, and the velocity and pressure fields surrounding the contrast microbubble, to better understand sonoporation.

## 1. Introduction

Ultrasound waves are pressure waves capable of transporting energy into the body at precise locations as they are absorbed relatively little by tissues. Their non-invasive, safe and painless transmission through the skin make them to be widely used in drug delivery and gene therapy applications through cavitating gas bodies such as microbubbles (Pitt, Husseini et al. 2004).

Microbubbles can carry and transport drugs/genes on or within their shells to the desired site within the body. Physicochemical interactions can be used to increase the specificity of microbubbles targeting while limiting the toxicity of the drug to healthy tissues. There are several ways of targeting the drug carrying microbubble to the desired tissue like cationic charging the surface of the bubble or attaching ligands to the bubble surface that can bind to the receptors of cells (Ferrara, Pollard et al. 2007). When the drug carrying microbubble reaches the desired site, the ultrasound wave can disrupt the microbubble and release the drug.

In addition to the drug carrier role, microbubbles in the presence of ultrasound facilitates the transportation of drugs, whether free or bounded to a drug carrier through a process called sonoporation. Sonoporation is the acoustically induced temporary rupture of membranes. Ultrasound waves can also spontaneously contract and expand dissolved gas and vaporized liquid, forming cavitation gas bubbles in tissue (Liang, Tang et al. 2010). With a scanning electron microscope, it has been shown that multiple surface pores are created on the cell membrane exposed to ultrasound (Tachibana, Uchida et al. 1999). The mean radius of single pores on cell membrane can reach 110 nm in the presence of Definity microbubbles with ultrasound (0.2 s, 0.3 MPa, 1.075 MHz) (Zhou, Kumon et al. 2009). The induction of sonoporation using targeted microbubbles have been studied both in vitro and vivo (Kooiman, Foppen-Harteveld et al. 2011, Escoffre, Zeghimi et al. 2013, Lee, Yoon et al. 2016). It has been shown that targeted microbubbles can induce sonoporation of endothelial cells in vivo, thereby making it possible to combine molecular imaging and drug delivery (Skachkov, Luan et al. 2014).  The targeted microbubbles are contrast agents as they were initially developed for enhancing the contrast of the image inside body due to high echogenicity. Contrast agents consist of a gas core encapsulated with a layer of protein, lipid, and phospholipid to prevent them against early dissolution. The encapsulation reduces the surface tension and therefore reduces the diffusion of gas inside the microbubble into the medium. Contrast agent in the presence of ultrasound may show two different kinds of motions, stable cavitation and inertial cavitation. Stable cavitation occurs at low excitation pressures making the microbubbles oscillate repeatedly over many cycles. The stable cavitation can create circulation fluid flow around the microbubble (microstreaming) capable of rupturing the cells and vesicles due to high velocities and high shear rates (Marmottant and Hilgenfeldt 2003). Inertial cavitation occurs at high amplitude ultrasound excitation pressures causing the bubble to implode. If the collapsing microbubbles are in the vicinity of a solid boundary, they may generate a high velocity liquid microjet penetrating toward the wall. The impingement of microjet along with its high velocity may create holes, and high shear stresses in cell membrane which may result in sonoporation which consequently facilitates the uptake of drugs into the desired site (Lokhandwalla and Sturtevant 2001, Sundaram, Mellein et al. 2003, Newman and Bettinger 2007).  The study of non-spherical bubble collapse shows that the dynamic of the jet strongly depends on the distance of the bubble relative to the wall (Calvisi, Iloreta et al. 2008). It is also shown that in addition to the jet speed, jet size can have a determining factor on the impacting power of collapsing bubble (Fong, Klaseboer et al. 2008). During the formation of jet, there exists very high pressure fluid on top of the jet pushing the bubble toward the wall (Shervani-Tabar and Mobadersany 2013, Shervani-Tabar and Mobadersany 2013).  The micorojet impact on the cell membrane has an important role in sonoporation, and therefore in this paper we do a numerical study on the formation of microjet from contrast microbubbles in the vicinity of a cell membrane to better understand the mechanism of sonoporation.

We have assumed the membrane as a rigid wall (the elastic modulus of cell membranes is estimated to be on the order of 10–1000 MPa; i.e., the elastic modulus of the membrane of lymphocytes is 80 MPa. such high values show that a cell is a rather rigid object(Doinikov and Bouakaz 2010)). We have used interfacial rheology models with zero thickness to simulate the encapsulation of the contrast microbubbles (Chatterjee and Sarkar 2003). The boundary element method has been used for the numerical study. Many researches have been conducted to study the formation of jets in bubbles using boundary element method. Axisymmetric jetting for acoustic bubbles in an infinite liquid was studied by (Calvisi, Lindau et al. 2007, Wang and Blake 2010, Wang and Blake 2011), and near a boundary subjected to ultrasound by (Klaseboer and Khoo 2004, Klaseboer and Khoo 2004, Fong, Klaseboer et al. 2006, Curtiss, Leppinen et al. 2013, Hsiao, Choi et al. 2013, Blake, Leppinen et al. 2015). The simulation of the dynamics of uncoated and coated microbubbles near a wall under high intensity traveling ultrasound wave show that the jet direction depends on the propagation direction of the wave and the distance of the microbubble from the wall (Wang and Manmi 2014, Wang, Manmi et al. 2015). Wang, et al. applied modified Rayleigh-Plesset equation to simulate coating behaviour stating the bubble is spherical most of its life time. Hao Yu, et al. studied dynamics of the microbubble and cell membrane assuming the cell membrane as a fluid-fluid interface. They have assumed the surface tension of each element on the contrast microbubble is equal during the simulation (Yu, Lin et al. 2015). In this study we simulated the collapse of contrast agents using interfacial rheology models for modeling the encapsulation. These models were originally developed for spherical oscillations assuming surface tension is the function of elasticity and bubble size. We have modified the encapsulation models as the function of elasticity and surface area of each element on the microbubble since the bubble is not spherical during the collapse phase. We have also studied the effect of different parameters on the dynamics of the contrast agent, the surrounding velocity and pressure fields, and the induced shear stress on the wall due to the its collapse to better understand sonoporation.

## 2. Mathematical equations

In this study, the flow around the contrast microbubble (contrast agent) is assumed incompressible, irrotatational and inviscid. Therefore the surrounding flow is governed by Laplace equation. So, the boundary element method for potential flow is applied to study the contrast agent using the following Green's integral formula.

$$c(p)\phi(p) + \int_S \phi(q) \left\{ \frac{\partial}{\partial n} \left( \frac{1}{|p-q|} + \frac{1}{|p-q^{\wedge}|} \right) \right\} ds$$
$$= \int_S \frac{\partial \phi(q)}{\partial n} \left( \frac{1}{|p-q|} + \frac{1}{|p-q^{\wedge}|} \right) ds \ .$$
(1)

In equation(1), $s$ is the boundary of the fluid domain which includes the bubble surface and its image; $\phi$ is the velocity potential and $\partial\phi/\partial n$ is the normal velocity; $p$ is any point in the fluid domain, or on its boundary; q is any point on the fluid boundary and $q^{\wedge}$ is the image point of q in plane z=0 where the cell membrane is located. $c(p)$ is a coefficient dependent on the location of the point p (for $p$ on the bubble $c(p) = 2\pi$, and for $p$ inside the fluid $c(p) = 4\pi$). The image bubble satisfies the impermeability condition on the cell membrane.

Figure 1 shows the schematic of the problem. The bubble surface is discretized into M cubic spline elements. The velocity potential and the normal velocity are assumed to be constant on each element located in the middle of the elements.

The normal velocity on the boundary of the fluid domain is indicated by n and is directed outward the fluid (directed toward bubble center). The vertical axis is indicated by z, while r is the radial axis. The problem is assumed axisymmetric, and therefore $\phi$ and $\partial\phi/\partial n$ are independent of azimuth angle.

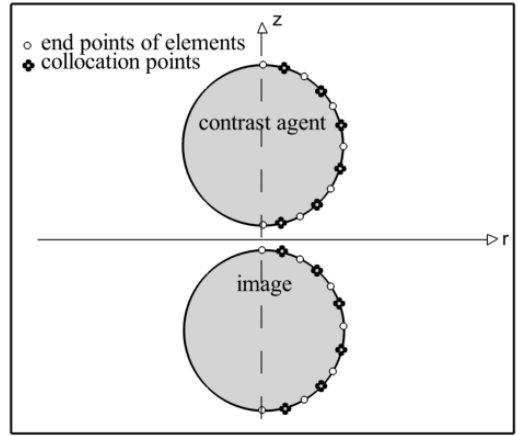

**Figure 1** Schematic of the problem

Equation (1) gives a set of linear equations for evaluating $\partial \phi / \partial n$ on each segment on microbubble surface. The evaluation of the elements of the integrands were calculated numerically using Gauss-Legendre quadrature, unless the collocation point was within the segment making the integrand to be singular which was treated specially (for more details see (Taib 1985)).

In this research, we are studying Sonazoid as our desired contrast microbubble consisted of a perfluorocarbon gas core and a lipid shell as an encapsulation. The cell membrane is assumed as a rigid surface. By having distribution of the velocity potential on the bubble surface, tangential velocity on the bubble surface is obtained by differentiation of the velocity potential along the surface of the bubble. Thus, by having the velocity potentials on the collocation points on the bubble surface, the normal and tangential velocities on the surface of the bubble are evaluated and consequently the microbubble shape at the next time step is obtained using the second order Runge-Kutta scheme. Similarly the velocity potential is being updated using the second order Runge-Kutta scheme. It is worth mentioning that during the evolution of the contrast microbubble, instability occurs which is removed by employing a five point smoothing technique as shown in the following (Pearson, Blake et al. 2004). The instability only occurs during the collapse of the bubble when the jet forms.

$$\bar{f}_i = f_i - \frac{1}{16}\left(f_{i-2} - 4 f_{i-1} + 6 f_i - 4 f_{i+1} + f_{i+2}\right) . \tag{2}$$

$\bar{f}_i$ is the smoothed quantity of $f_i$ which can be r, z and $\phi$. As discussed above, to solve the set of linear equations given in(1), the value of the velocity potential on the bubble surface should be known at each time step. As mentioned earlier, the initial velocity potential on the bubble surface is zero. To find the velocity potential in the next time steps, the unsteady Bernoulli equation is applied for the flow along the surface of the contrast agent.

$$\rho\left(\frac{D\phi}{Dt} - \frac{1}{2}|\nabla \phi|^2\right) + \rho g(z-h) = P_\infty - P_{bw} . \tag{3}$$

$\rho$ is the fluid density surrounding the microbubble, $\phi$ and $\nabla \phi$ are the velocity potential and the velocity on the bubble surface respectively. $P_{bw}$ is the pressure of the fluid on the bubble surface.

$P_\infty$ is the pressure in the far field including the static ambient pressure (here atmospheric pressure) and excitation ultrasound pressure.

$$P_\infty = P_{atm} - P_{ex}\sin(\omega t). \qquad (4)$$

$P_{ex}$ and $\omega$ are the excitation pressure and excitation radial frequency due to ultrasound wave respectively. Note that the buoyancy force is neglected because of the micron size of contrast agents. The problem under investigation is non-dimensionalized by employing the initial radius of the microbubble $R_0$, the liquid density $\rho$, and the atmospheric pressure $p_{atm}$. Normalized form of Equation(3) is given as:

$$\frac{D\phi^*}{Dt^*} = \frac{1}{2}\left|\nabla^*\phi^*\right|^2 + \frac{(P_\infty - P_{bw})}{P_{atm}}. \qquad (5)$$

$\phi^*$, $t^*$ and $\nabla^*$ are the non-dimensional velocity potential, time and gradient respectively. Therefore the non-dimensional velocity potential on the next time step is calculated using the following forward Euler method.

$$\phi^{*n+1}_j = \phi^{*n}_j + \left(\frac{D\phi^*}{Dt^*}\right)^n_j \times \Delta t^*. \qquad (6)$$

$\Delta t^*$ is the desired non-dimensional time step obtained from Bernoulli equation.

$$\Delta t^* = \frac{\Delta\phi^*}{\max\left(\frac{P_\infty - P_{bw}}{p_{atm}} + \frac{1}{2}|\nabla\phi|^2\right)}. \qquad (7)$$

In equation(7), $\Delta\phi^*$ is some constant that controls the maximum increment of the velocity potential on the bubble surface between two successive time steps.

The gas inside the contrast agent is assumed to be an ideal gas following polytropic behaviour.

$$P_g = P_{g0}\left(\frac{V_0}{V}\right)^\kappa. \qquad (8)$$

$P_g$ is the gas pressure inside the contrast microbubble when its volume is $V$. $P_{g0}$ is the initial gas pressure inside the contrast microbubble when it is in its initial volume $V_0$, and $\kappa$ is the polytropic constant. For the Sonazoid contrast agent, continuity of the heat flux at the bubble interface shows that temperature drop basically occurs inside the bubble. It has shown that the temperature difference between the bubble surface and the inside gas is much more than the temperature difference between bubble surface and the surrounding liquid. (Kamath, Prosperetti et al. 1993, Brenner, Hilgenfeldt et al. 2002). This results in the assumption of adiabatic process to the gas inside Sonazoid.

### 3. Modelling the encapsulation of the microbubble

The pressure inside and outside the microbubble are related by the normal stress balance. The effect of the encapsulation must be considered on the normal stress balance. The encapulation of the microbubble is assumed as a viscoelastic interface having elastic surface tension and dilatational viscosity.

$$P_g - P_{bw} = \left(\sigma_{eff}(R) + \kappa^s(\nabla_s \cdot V)\right)(\nabla_s \cdot n). \tag{9}$$

In equation(9), $\sigma_{eff}$ and $\kappa^s$ are the effective surface tension and dilatation viscosity of the contrast agent due to the encapsulation respectively, $\nabla_s \cdot V$ is the surface divergence of velocity, and $\nabla_s \cdot n$ is the curvature of the contrast agent(Scriven 1960). Applying the dilatation viscosity term in the normal jump condition gives rise to instability which could not be removed by smoothing. Therefore in the current research, we neglect the dilatation viscosity for simplicity, and we will study its effect in future works. The curvature of microbubble is as follows:

$$\begin{aligned}\nabla_s \cdot n =& \left(\frac{dz}{d\xi}\frac{d^2r}{d\xi^2} - \frac{dr}{d\xi}\frac{d^2z}{d\xi^2}\right) \bigg/ \left(\left(\frac{dr}{d\xi}\right)^2 + \left(\frac{dz}{d\xi}\right)^2\right)^{3/2} \\ & - \left(\frac{dz}{d\xi}\right) \bigg/ \left(r(\xi)\left(\left(\frac{dr}{d\xi}\right)^2 + \left(\frac{dz}{d\xi}\right)^2\right)^{1/2}\right).\end{aligned} \tag{10}$$

$\xi$ is the arc length parameter describing r and z. The effective surface tension $\sigma_{eff}(R)$ describes the interfacial rheology of the encapsulation. For a free bubble, surface tension $\sigma_w$ is at a pure air-water interface.

Several models have been proposed to study the interfacial rheology, i.e. $\sigma_{eff}(R)$. In the present study we have used two different models to calculate the effective surface tension.

### 3.1. Viscoelastic model with exponentially varying elasticity (EEM) (Paul, Katiyar et al. 2010)

$$\begin{cases} \sigma_{eff}(R) = \sigma_0 + \beta E^s, & \text{where } E^s = E_0^s \exp(-\alpha^s \beta) \text{ and } \beta = (\frac{R^2}{R_E^2} - 1), \\ \sigma_{eff}(R) = 0, & \text{for} \quad \sigma(R) < 0. \end{cases} \tag{11}$$

$\sigma_{eff}(R)$ is the effective interfacial tension, $E^s$ is the elasticity of the shell, $E_0^s$ is elasticity constant and $\beta$ is the area fraction. $R_E$ is the equilibrium radius of the contrast agent. Equilibrium radius equals to the radius of the bubble where elastic stress is zero.

$$R_E = R_0[1 + (\frac{1 - \sqrt{1 + 4\gamma_0 \alpha^s / E_0^s}}{2\alpha^s})]^{-1/2}. \tag{12}$$

$R_0$ is the initial radius of contrast microbubble. This model has an upper limit on the effective interfacial tension. We also set the lower limit of the effective surface tension to be zero in this study. To apply the EEM model to our boundary element study, we substitute the area fraction with the following equation:

$$\begin{cases} \beta = (\dfrac{R^2}{R_E^2}-1) = (\dfrac{A}{A_E}-1) , \\ A_E = A_0[1+(\dfrac{1-\sqrt{1+4\sigma_0\alpha^s/E_0^s}}{2\alpha^s})]^{-1} . \end{cases} \quad (13)$$

$A$, $A_0$ and $A_E$ are the surface area, initial surface area and equilibrium surface area of each element on the contrast agent.

### 3.2. **Marmottant model (MM)** (Marmottant, van der Meer et al. 2005)

$$\sigma_{eff}(R) = \begin{cases} 0 & for \quad R \leq R_{buckling} , \\ \chi\left(\dfrac{R}{R_{buckling}}-1\right) & for \quad R_{buckling} \leq R \leq R_{rupture} , \\ \sigma_w = 0.072 & for \quad R \geq R_{rupture} . \end{cases} \quad (14)$$

Where $R_{rupture} = R_{buckling}\left(1+\dfrac{\sigma_w}{\chi}\right)^{\frac{1}{2}}$ and $\chi$ is the elastic compression modulus. In this study the buckling surface area $R_{buckling}$ is chosen to be the initial radius of the contrast microbubble. Below the buckled radius $R_{buckliung}$, the microbubble is in the buckled state where the effective surface tension is zero, and above the rupture radius $R_{rupture}$, the interface acts as a pure air-water interface. To apply the Marmottant model to our boundary element study, we substitute the area fraction and rupture radius with the following equation.

$$\dfrac{R^2}{R_{buckling}^2} = \dfrac{A}{A_{buckling}} ,$$
$$A_{rupture} = A_{buckling}\left(1+\dfrac{0.072}{\chi}\right) . \quad (15)$$

$A$, $A_{rupture}$ and $A_{buckling}$ are the surface area, rupture area and buckling area of each element on the contrast agent. The buckled area is assumed as the initial surface area of the contrast agent. Table 1 shows the characteristic properties of Sonazoid according to different encapsulation models (Katiyar and Sarkar 2012).

**Table 1** Characteristic properties of Sonazoid contrast agent according to different encapsulation models

| **Exponential elasticity model (EEM)** (Paul, Katiyar et al. 2010) | $\sigma_0 = 0.019\,N/m, E_0^s = 0.55\,N/m, \kappa_s = 1.2e^{-8}Ns/m$ <br> $\alpha^s = 1.5, \kappa = 1.07$ |
|---|---|
| **Marmottant model** (Marmottant, van der Meer et al. 2005) | $\chi = 0.55\,N/m, \kappa = 1.07, \kappa_s = 1.2e^{-8}Ns/m$ |

## 4. Normalising the problem

As stated above the problem under investigation is normalized by employing the initial radius of the bubble $R_0$, the liquid density $\rho$, and the atmospheric pressure $p_{atm}$. The non-dimensional parameters are shown in table 2.

**Table 2** Non-dimensionalising the parameters

| | |
|---|---|
| $r^* = r/R_0$ | $\sigma_{eff}^* = \sigma_{eff}/(R_0 P_{atm})$ |
| $z^* = z/R_0$ | $P_g^* = P_g / P_{atm}$ |
| $h^* = h/R_0$ | $P_{ex}^* = P_{ex} / P_{atm}$ |
| $\phi^* = (\phi/R_0)\sqrt{\rho/P_{atm}}$ | $t^* = t/\left(R_0\sqrt{\rho/P_{atm}}\right)$ |
| $\nabla^* = \nabla \times R_0$ | $\partial\phi^*/\partial n^* = (\partial\phi/\partial n)\sqrt{\rho/P_{atm}}$ |
| $\omega^* = \omega R_0 \sqrt{\rho/P_{atm}}$ | $\tau^* = \tau / P_{atm}$ |

## 5. Validation

To validate the numerical results, we consider the oscillation of the contrast agent using boundary element method (BEM) in an infinite fluid in the absence of cell membrane where the contrast agent oscillates spherically under the excitation of ultrasound wave. We compare our results with the radial motion of the contrast agent obtained from the modified Rayleigh-Plesset (R-P) equation(16). Note that we have neglected the fluid viscosity in R-P equation to compare it with the solution of Laplace equation in BEM.

$$\rho\left(R\ddot{R} + \frac{3}{2}\dot{R}^2\right) = P_{g_0}\left(\frac{R_0}{R}\right)^{3\kappa} - \frac{2}{R}\sigma_{eff}(R) - P_{atm} + P_A \sin(\omega t). \tag{16}$$

Figure 2 shows the comparison of non-dimensional volume oscillation of the contrast agent (volume is non-dimensionlised by initial bubble radius) with respect to non-dimensional time between BEM and R-P while using EEM model for the encapsulation of contrast agent. The contrast agent has been excited by an ultrasound wave with $P_{ex} = P_{atm}$ and $f$ =2MHz.

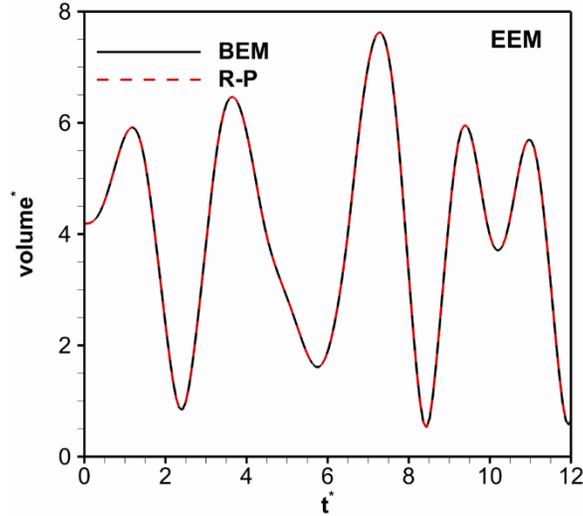

**Figure 2** Non-dimensional volume oscillation of contrast agent varying with non-dimensional time in an unbounded medium at $P_{ex} = 100 KPa, f = 2 MHz$ using EEM model for the encapsulation

Figures 3-4 show the collapse of the free bubble (bubble without encapsulation) and the comparison of numerical results with previous works for the case where the bubble near a rigid wall has been excited due to its initial high pressure. The strength of the initial high pressure is $\alpha = P_{g_0}/P_{atm} = 100$. It is seen that there is a very good agreement between the results. Figure 3 shows the evolution of the bubble at different non-dimensional times located initially below the rigid wall at $\gamma^* = -2$ in comparison with the results by (Best 1991). $\gamma^* = h/R_m$ is the standoff distance where h is the initial distance of the bubble center from the rigid wall and $R_m$ is the maximum radius of the bubble in an unbounded medium. In figures 3-4, maximum radius $R_m$ was used for normalising parameters instead of initial radius $R_0$.

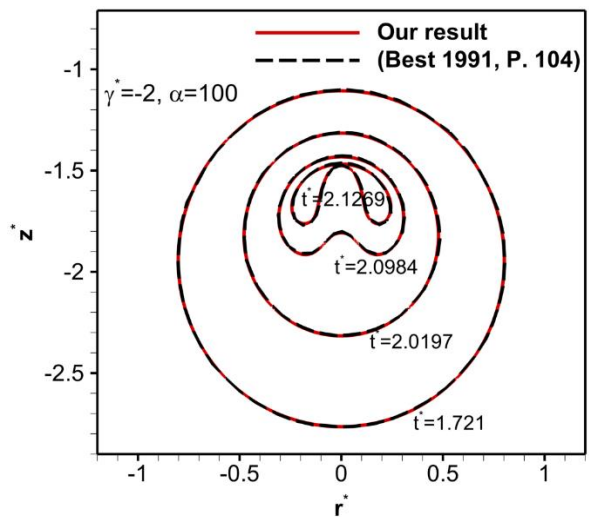

**Figure 3** The comparison of evolution of free bubble excited due to high initial pressure at different non-dimensional time when $\gamma^* = -2.0, \alpha = 100$ with (Best 1991)

Figure 4(a) shows the comparison of final stage of free bubble collapse when $\gamma^* = 1.5$, and figure 4(b) shows the comparison of non-dimensional volume at different standoff distances with the results obtained by Pearson et al. (Pearson, Blake et al. 2004).

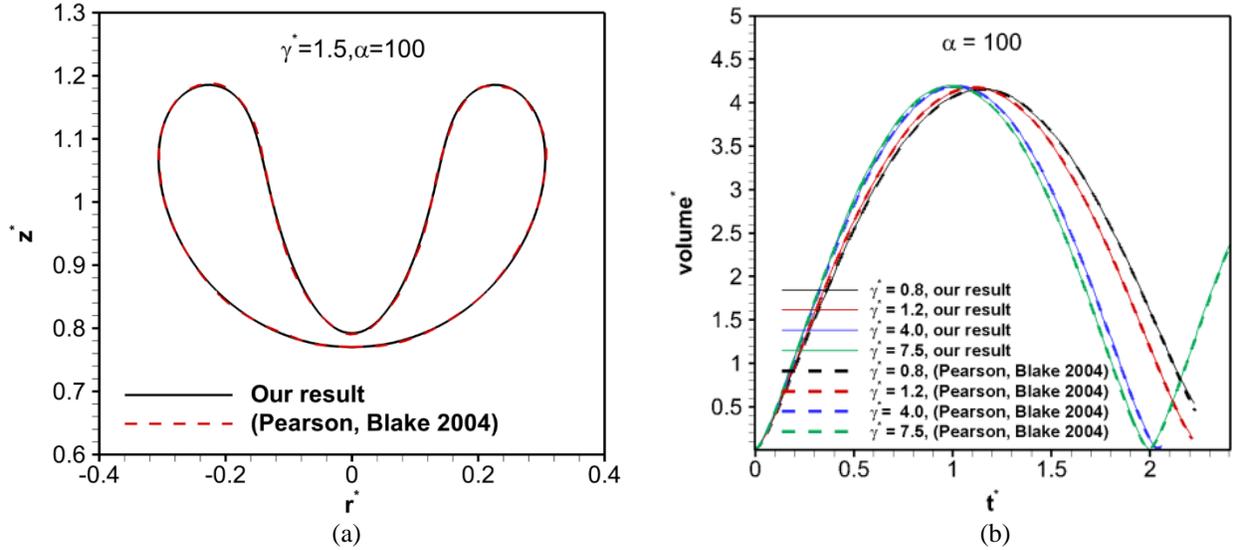

**Figure 4** (a) The comparison of final stage of free bubble collapse excited due to high initial pressure when $\gamma^* = 1.5$, $\alpha = 100$ with (Pearson, Blake et al. 2004), (b) The comparison of volume of free bubble excited due to high initial pressure when $\alpha = 100$ at different standoff distances with (Pearson, Blake et al. 2004)

## 6. Results and discussion

Sonoporation was demonstrated using ultrasound waves of different frequencies. The most commonly used frequencies are those in therapeutic (1 to 3 MHz) and diagnostic ultrasound transducers (3 to 18 MHz) (Jelenc, Jelenc et al. 2012). Therefore, in this study we are focusing on the MHz excitation frequency range. In this study, we are using Sonazoid as the desired contrast agent, and therefore we focus on the bubble radius size of $1.6\,\mu m$ which is the average size for Sonazoid. In this study, the excitation frequency is assumed to be 2 MHz which is the damped resonance frequency of Sonazoid at average size. At the desired frequency, exciting the contrast agent at low ultrasound pressures below the threshold level of inertial cavitation result in the stable cavitation. At the stable cavitation, the bubble oscillates repeatedly. Exciting the contrast agent at pressures above the threshold level of inertial cavitation makes the bubble collapse and form a jet penetrating toward the wall. The focus of this research is to study the behavior of contrast agent at high excitation pressures where the contrast agent collapses and forms a jet. In this study, the bubble has been discretized into ninety spline elements which is the optimum number of elements to get stable jet profiles in our case. Figure 5 shows the convergence of the evolution of the contrast agent using different time steps when the contrast agent is initially located at $h^* = 4.0$ and has been excited at $P_{ex}^* = 5.0$. Velocity potential and end points have been smoothed every ten time steps. Note that too much smoothing of the contrast agent may affect the results.

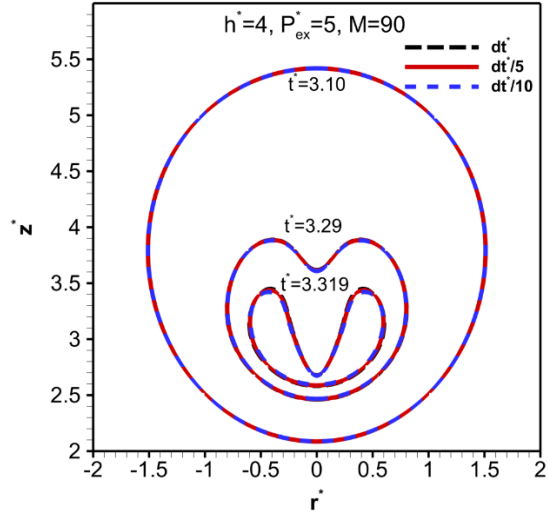

**Figure 5** Profile of the collapse of contrast agent using EEM for the encapsulation with different time steps when $h^* = 4.0, P_{ex}^* = 5.0$ with ninety elements

Before continuing to the result section, we want to compare our approach with the approach used by Wang, Manmi and Calvisi (Wang, Manmi et al. 2015) for the behaviour of contrast agent near a wall. Wang et al. have used modified Rayleigh-Plesset equation to simulate the behaviour of contrast agent replacing instant radius with inverse of local curvature assuming the bubble is spherical most of its life time while applying Church-Hoff model (Church 1995, Hoff 2001) for coating simulation. To compare the results, the shape of the contrast agent along with the jet tip velocity have been plotted in figure 6 using Church-Hoff model for the encapsulation in both methods. Church-Hoff model gives equation (17) to simulate the effective surface tension. Where $G_s$ and $d_{sho}$ are 0.52Mpa and 4nm respectively (characteristic properties of Sonazoid contrast agent).

$$\sigma_{eff}(R) = 6G_s d_{sho} \frac{R_0^2}{R^2}(1 - \frac{R_0}{R}) . \tag{17}$$

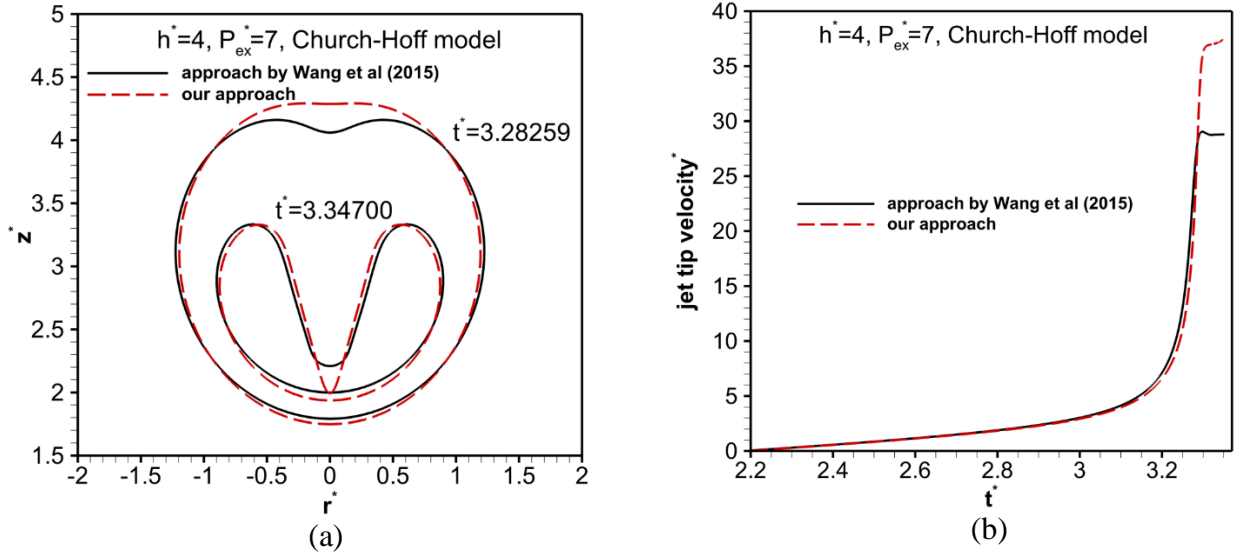

**Figure 6** (a) The comparison of final stage of contrast agent collapse and (b) jet tip velocity with non-dimensional time when $h^* = 4.0, P_{ex}^* = 7.0, f = 2MHz$ with the approach used by (Wang, Manmi et al. 2015)

As shown in figure 6(a) in our approach, the contrast agent starts forming a jet at $t^* = 3.28259$ while it has already formed a jet in the approach simulated by Wang et al. In our approach, it takes longer time for the contrast agent to form a jet. But despite the delay in forming a jet, it is seen that the jet progresses faster, and therefore figure 6(b) shows a higher jet tip velocity in our approach. To observe the behavior of contrast agent near a wall, figure 7 shows the contrast agent and the surrounding velocity and pressure when $h^* = 4.0, P_{ex}^* = 5.0$. $h^*$ is the non-dimensional initial distance of the bubble center from the wall and $P_{ex}^*$ is the non-dimensional excitation pressure. Note that the magnitude of velocity vectors in the figure is relative to other particles in the same figure, and they do not show the actual magnitude of fluid particles velocity. Figure 7(a) shows the contrast agent at the early stages of the growing phase where the bubble expels the surrounding fluid due to higher pressure inside. Figure 7(b) show the contrast agent when it reaches the maximum size at $t^* = 2.1888$ where $t^*$ shows the non-dimensional time. As it is observed the bubble expands nearly spherically to reach the maximum volume. It is shown that the fluid underneath the bubble is moving away from the bubble and is redirected toward the bubble at the middle. During the last stage of the collapse phase, the contrast agent forms a jet directed toward the wall. Figure 7(c) shows the contrast agent during the collapse phase when it starts to form a jet at $t^* = 3.2697$. As it is observed there exists a high-pressure region on top of the contrast agent resulting in the formation of jet. This high pressure region pushes the contrast agent to form a high velocity jet like what was observed for a bubble driven with a high initial pressure (Shervani-Tabar, Mobadersany et al. 2011). Figure 7(d) shows the last stage of the collapse phase of the contrast agent at $t^* = 3.3214$. As it is shown the collapse time from the start of the jet formation to the end of the collapse is very short.

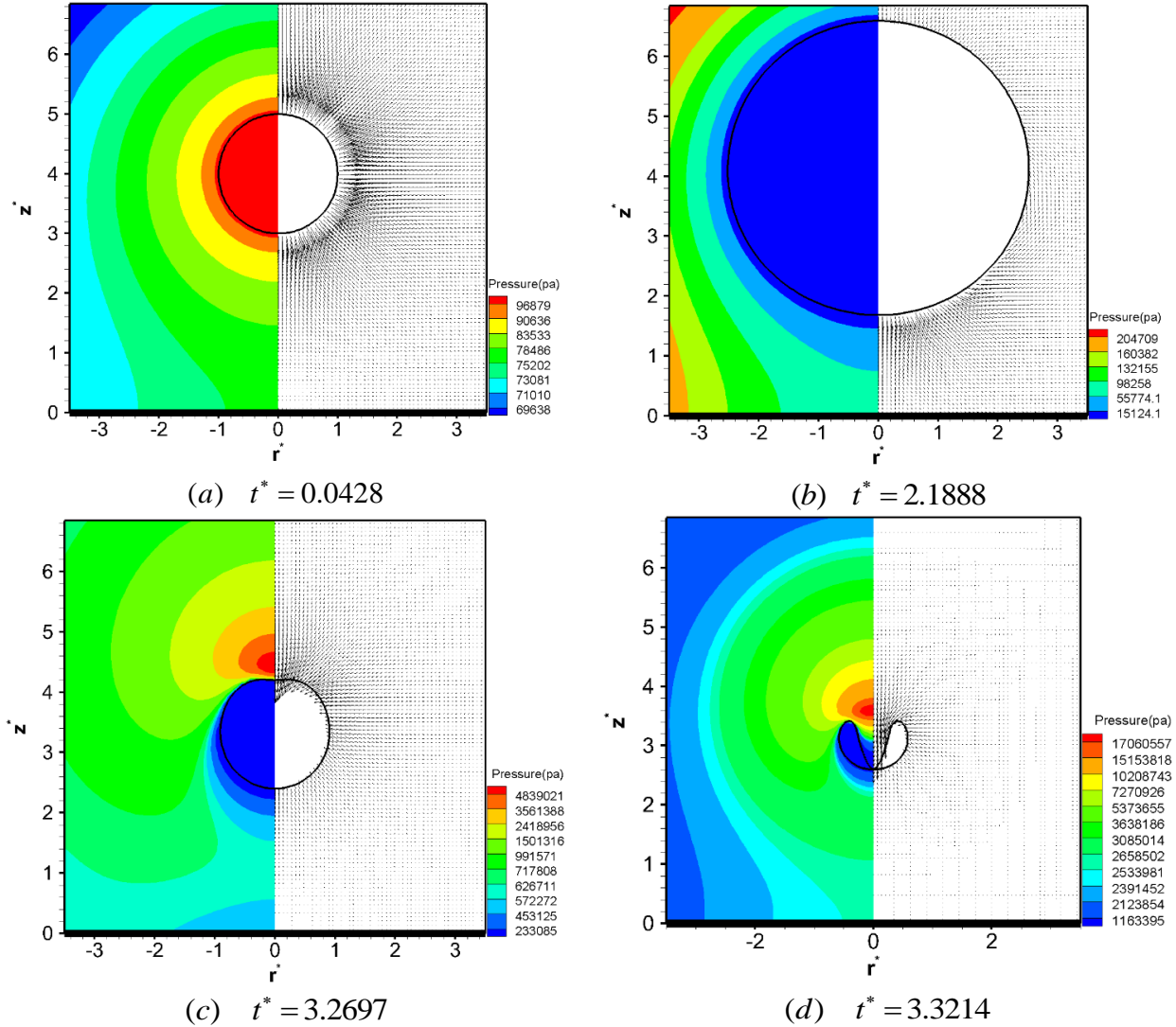

**Figure 7** Velocity and pressure surrounding the contrast agent with respect to non-dimensional time using EEM model for the bubble encapsulation when $h^* = 4.0, P_{ex}^* = 5.0$

As it is observed in figure 7(c-d), the microjet of the contrast agent and the adjacent surrounding fluid are moving toward the cell membrane with a very high velocity. This high velocity fluid impinges the wall and spreads radially along it. This will generate high velocity gradient on the wall, and therefore it will generate shear stress resulting in the rupture and perforation of the wall. To calculate the shear stress generated on the wall, we just consider the flow field induced from the jet of the contrast agent spreading outward radially along the wall (Ohl, Arora et al. 2006). Therefore, we assume a jet flow impacting vertically on the wall with a constant velocity equal to the velocity of the jet of the contrast agent. This velocity is the averaged vertical velocity of the elements on the jet at each time. Glauret wall jet flow has been used to estimate the shear on the wall (Glauert 1956). Glauert obtained this solution through a similarity analysis. In his derivation and analytical solution of the boundary layer equation, he had to neglect the region around the stagnation point (i.e., near to the point of impact). Glauert's similarity solution for the wall shear stress $\tau$ is in the following:

$$\tau = \rho v \left(\frac{\partial u}{\partial y}\right)_{y=0} = \rho \left(\frac{125 F^3}{216 v x^{11}}\right)^{1/4}. \qquad (18)$$

The constant F is the momentum flux of the incoming jet. For a jet impinging with a flat velocity profile, the momentum flux is shown in equation (18), where $U_{jet}$ is the averaged jet velocity, and $d_{jet}$ is the diameter of the jet of the contrast agent at each time.

$$F = \frac{1}{128} U_{jet}^3 d_{jet}^4 . \qquad (19)$$

From equations (18)(19) it is shown that the shear stress exerted on the wall has a direct relation with the jet velocity and jet width.

Figures 8 shows the average shear stress along the wall due to the impact of jet in figure 7. Equation (18) along with (19) have been used to obtain the average shear stress on the wall.

$$\tau_{average} = \sum \tau.(\Delta t)/t_{jet} . \qquad (20)$$

$t_{jet}$ is the total time of the jet formation, and $dt$ is the variable time step.

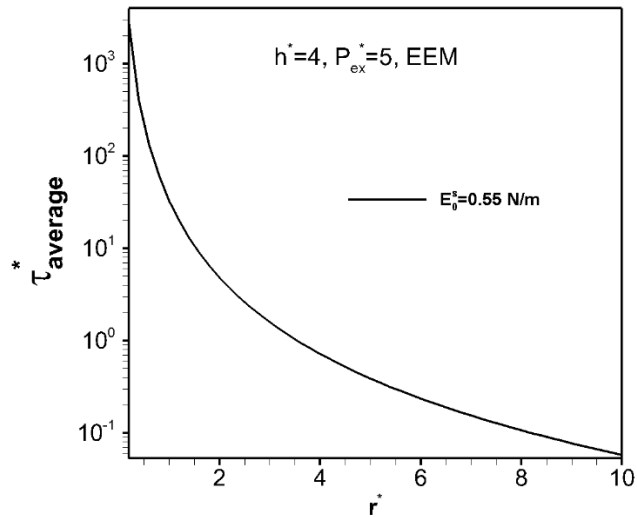

**Figure 8** Corresponding shear stress on the rigid boundary with respect to non-dimensional time when $h^* = 4, P_{ex}^* = 5$ using EEM model for the encapsulation

As mentioned earlier, the contrast agent forms a jet when excited at pressures above the threshold level of inertial cavitation. The threshold level not only depends on the amplitude of excitation pressure, but also depends on the initial distance of the contrast agent from the wall $h^*$. Figure 9 shows that depending on the initial distance of the contrast agent from the wall $h^*$, the contrast agent would form a jet at excitation pressures beyond line 2. Figure 9 has been plotted using EEM model assuming constant elasticity to be $E_0^s = 0.55 N/m$. It is shown that higher excitation pressure is required to form a jet at higher initial distances from the wall. To compare the effect of encapsulation elasticity on jet formation, lines 1 and 3 shows the minimum excitation pressure level where the contrast agent forms a jet at constant elasticities of 0.2 N/m and 0.9 N/m respectively. It is shown that the minimum excitation pressure where the contrast agent can form a jet at the desired initial distance from the wall is increasing with the increase of elasticity.

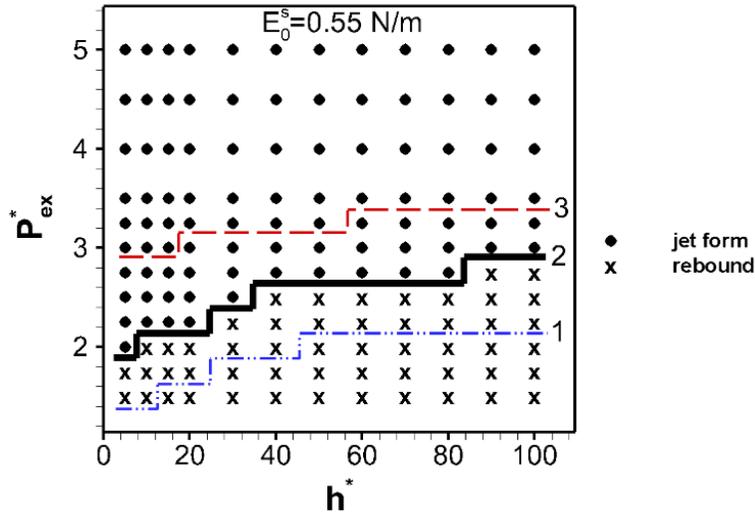

**Figure 9** Excitation pressure of the contrast agent with respect to non-dimensional initial distance using EEM for the bubble encapsulation when $E_0^s = 0.55\,N/m$

Figure 10(a-d) shows the velocity and pressure around the contrast agent with different elasticity at the last stage of the collapse phase excited with $P_{ex}^* = 5.0$ when EEM has been used for the encapsulation. As it is seen the fluid velocity is very high near the jet of the bubble. Also, it is seen that there exists a high-pressure region in the fluid on top of the jet. This high-pressure region is higher when contrast agent has higher encapsulation elasticity. This is evident from the lower volume of the contrast agent when elasticity of the encapsulation is higher. The fluid pressure on top of the bubble must be much higher than the pressure inside to be able to deform the bubble non-symmetrically. Since the lower volume of the bubble indicates the higher pressure inside the bubble, higher fluid pressure is needed to deform the bubble which has lower volume. Therefore, for contrast agent with higher elasticity, the fluid pressure on top of the bubble should be higher to overcome the high pressure inside the bubble to form a jet.

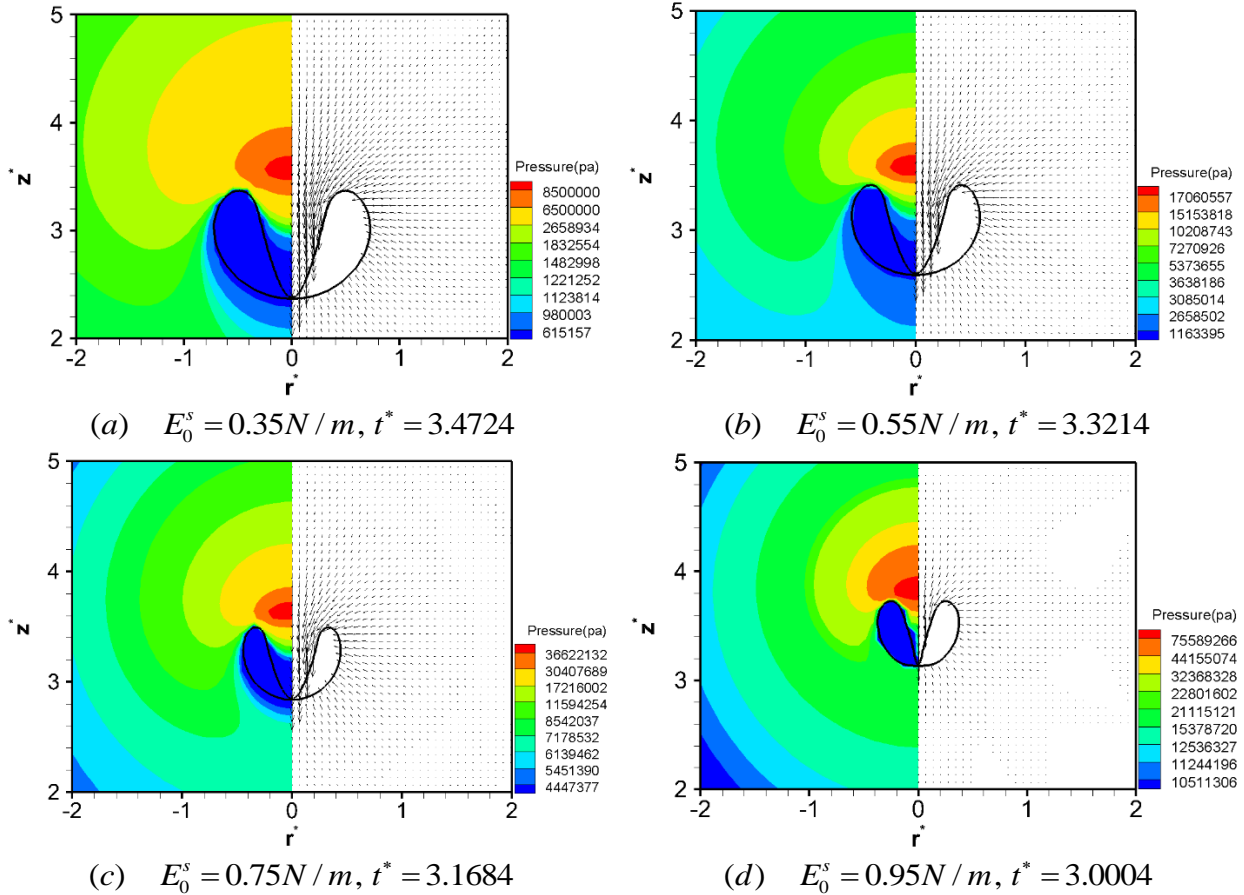

(a)  $E_0^s = 0.35 N/m, t^* = 3.4724$

(b)  $E_0^s = 0.55 N/m, t^* = 3.3214$

(c)  $E_0^s = 0.75 N/m, t^* = 3.1684$

(d)  $E_0^s = 0.95 N/m, t^* = 3.0004$

**Figure 10** Velocity and pressure surrounding the contrast agent using EEM model for the encapsulation with different shell elasticity constants when $h^* = 4.0, P_{ex}^* = 5.0$

Figure 11 (a-c) show the non-dimensional volume, volume centroid and jet tip velocity of the contrast agent at the conditions shown in figure 10 using EEM model for the encapsulation. Figures 11 (d-f) show the same results using MM for the encapsulation for comparison. As shown in figures 11(a) the maximum growth of the contrast agent decreases with the increase of elasticity, as the higher elasticity makes the contrast agent stiffer. Figure 11(b) shows the volume centroid of the contrast agent moving away from the membrane during the bubble expansion and moving toward the membrane during the bubble collapse. The similar behavior for the volume centroid has been shown for thick-shelled contrast agents subjected to pressure waves (Hsiao, Lu et al. 2010). It is also seen that the contrast agent with higher elasticity tends to have less movement toward the cell membrane which is expected by looking at the life time of the contrast agent in figure 11(a). It is shown that the life time of the contrast agent is decreasing by the increase of elasticity. Therefore, contrast agent with higher shell elasticity, has less time to move toward the cell membrane.

Figure 11(c) shows the velocity of the jet tip at the axisymmetric axis with respect to non-dimensional time. As it is observed the bubble with higher elasticity tends to have larger jet velocity especially when using EEM for the encapsulation, this can be inferred by the high fluid pressure region shown in figure 10. For contrast agent with higher shell elasticity, the fluid pushes the bubble with stronger pressure to form a jet which results in higher jet velocity.

As it is seen from figures 11(d-f), the elasticity variation has a slight effect on the contrast agent behaviour using MM for the encapsulation. The comparison of contrast agent with free bubble volume change in figure 11, reveals that free bubble has much more expansion, more translation toward the cell membrane and less jet velocity in comparison with the contrast agent using both encapsulation models.

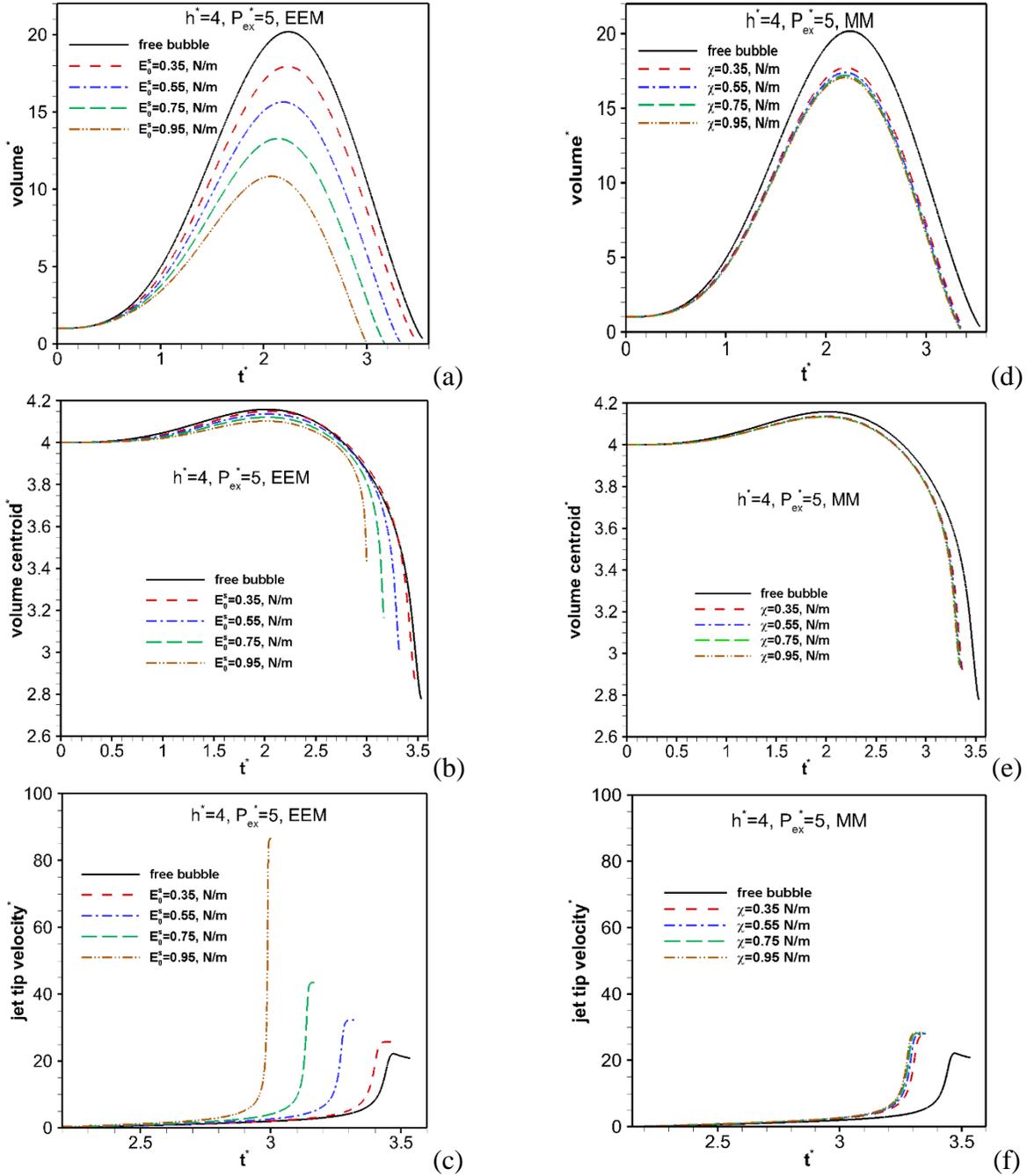

**Figure 11** Non-dimensional volume, non-dimensional volume centroid and non-dimensional jet tip velocity of microbubble with respect to non-dimensional time when $h^* = 4.0, P_{ex}^* = 5.0$ (a-c) using EEM model for the encapsulation (d-f) using MM for the encapsulation

To better understand the sensitivity of elasticity variation on contrast agent using the above-mentioned encapsulation models, figure 12 shows the effective interfacial tension of the first element of the contrast agent (element at the top near the axis of symmetry). It is shown that the effective interfacial tension is not changing significantly using Marmottant model since the model does not let the effective interfacial tension to go beyond the surface tension of free bubble.

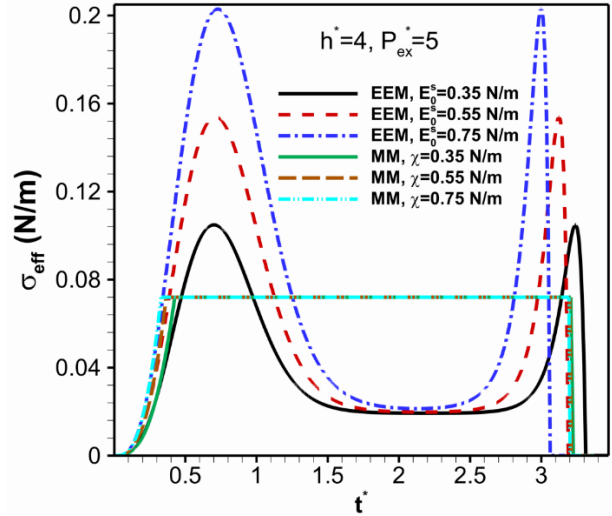

**Figure 12** Effective surface tension of the first element of the contrast agent (element at the top near the axes of symmetry) with respect to non-dimensional time when $h^* = 4.0, P_{ex}^* = 5.0$ using both EEM and MM for the encapsulation

Figure 13 shows the average jet radius and average jet velocity of the contrast agent over the total time of jet formation at different excitation pressures with different encapsulation elasticities using both EEM and MM model. The contrast agent is initially located at $h^* = 4$, and is excited at $f = 2MHz$. As it is shown in figure 13(a), the width of the jet is decreasing nearly linearly with the increase of elasticity using EEM model. Elasticity resists the collapse of the contrast agent, and hence results in jet formation at lower volumes. Therefore, at higher shell elasticity, the jet forms when the bubble reaches very small volume which causes the width of the jet to be smaller as well. Also, it is worth mentioning that the contrast agent forms a sharper jet in comparison with the free bubble (the non-dimensional average jet radius of the free bubble is 0.4557 at $P_{ex}^* = 5, h^* = 4$). As it is seen the elasticity variation has significant effect on the jet formation when using EEM for encapsulation, but the effect is negligible when MM model has been used for the encapsulation simulation. Figure 13(b) shows that the average jet velocity of the contrast agent is increasing exponentially by the increase of shell elasticity at lower excitation pressures when using EEM model for encapsulation. At higher excitation pressures, the average jet velocity is increasing almost linearly with the increase of shell elasticity.

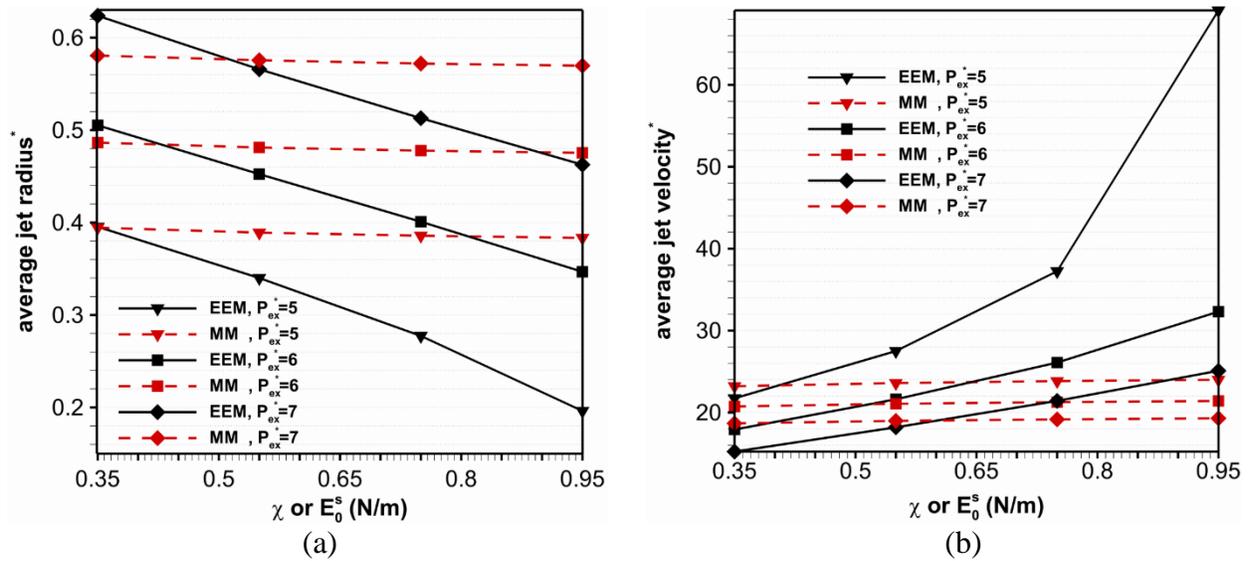

**Figure 13** (a) Non-dimensional average jet radius and (b) non-dimensional average jet velocity of the contrast agent with respect to different shell elasticity at different excitation pressures when $h^* = 4.0$

Figures 14 shows the shear stress on the wall due to the impact of jet with respect to non-dimensional time at two different distances away from the axis of symmetry using both EEM and MM for modelling the encapsulation of contrast agent. As mentioned earlier, the shear stress is induced by the jet on the wall, and the wall is assumed to be hit by the jet velocity. Therefore, figure 14 has shown the shear stress from the start of the jet to the end of bubble collapse.

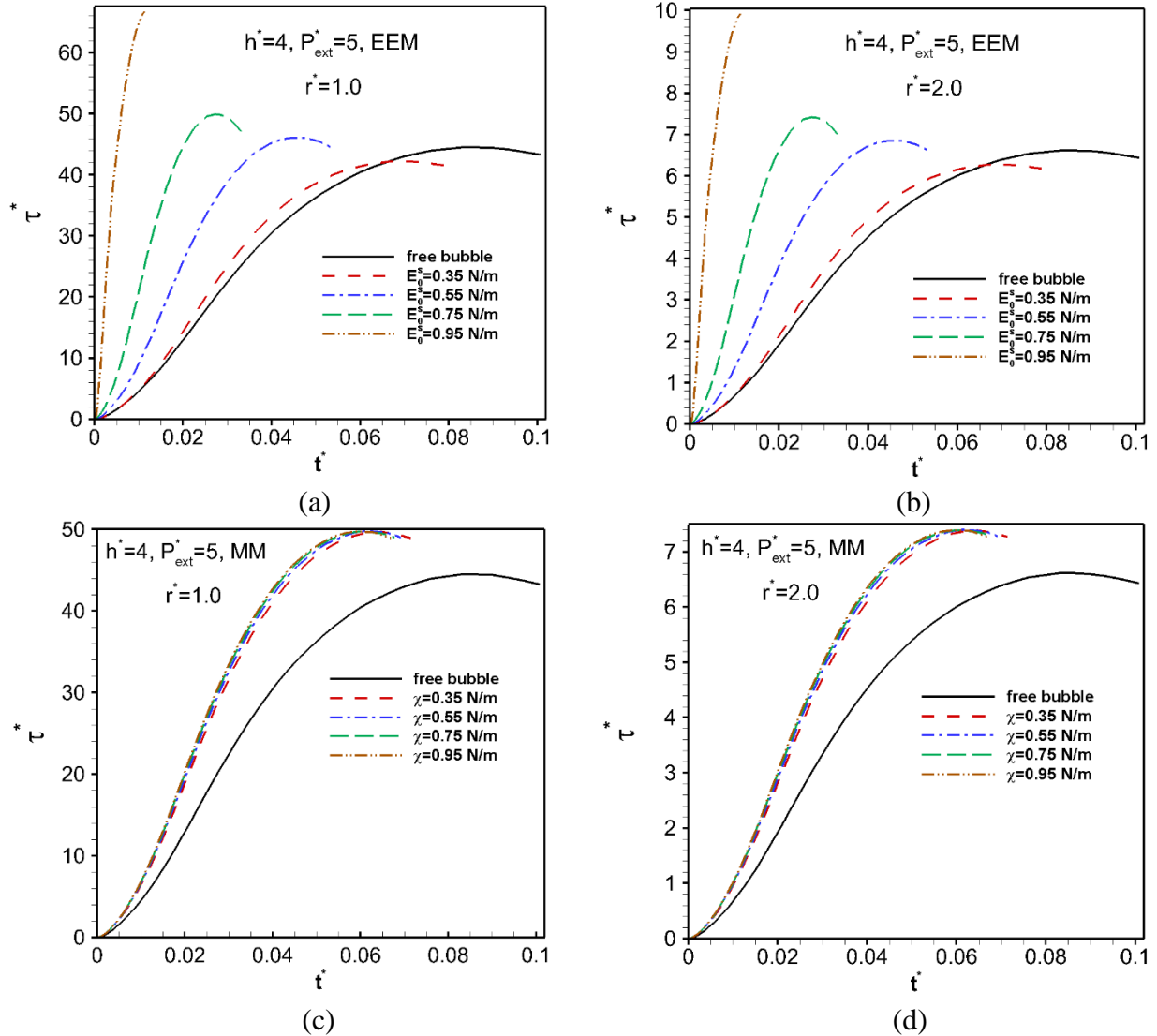

**Figure 14** Corresponding shear stress on the rigid boundary with respect to non-dimensional time at different elasticities of contrast agent when $h^* = 4.0, P_{ex}^* = 5.0,$ (a) at $r^* = 1.0$ using EEM model, (b) at $r^* = 2.0$ using EEM, (c) at $r^* = 1.0$ Using MM and (d) at $r^* = 2.0$ using MM for the encapsulation

Figure 14(a) and figure 14(b) show the shear stress on the wall at $r^* = 1.0, r^* = 2.0$ respectively using EEM for the encapsulation. As it is shown earlier in figure 8, and it is seen in figure 14(b), the shear stress decreases as we go far from the axis of symmetry. Also, as shown in figure 14(a-b), the shear stress on the wall is higher at higher elasticities. But the total time from the start of the jet to the end of the collapse is lower, and therefore the wall is under shear stress for a shorter time. Figure 14(c) and 14(d) show the shear stress on the wall at $r^* = 1.0, r^* = 2.0$ respectively using MM for the encapsulation. As it is seen the change in elasticity has no considerable effect on the shear stress.

Figure 15 show the resultant shear stress on the wall at different excitation pressures due to the jet impact using EEM for modelling the encapsulation. It is shown that the shear stress rises more rapidly at lower excitation pressures since the life time of the jet is shorter and the contrast agent reaches the last stage of the collapse earlier. But the shear stress at the last stage of the collapse is higher at higher ambient pressure.

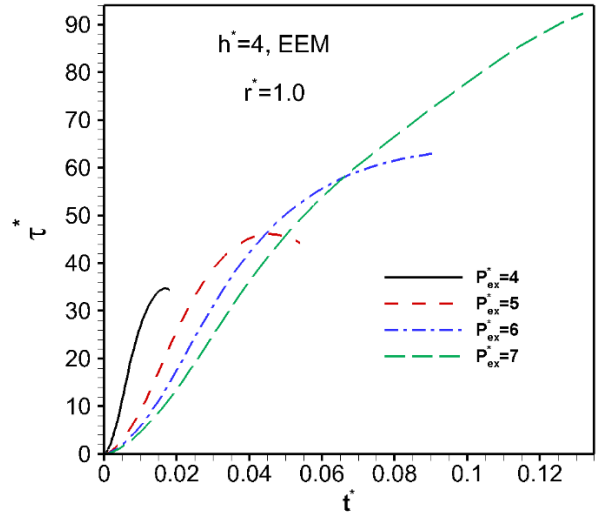

**Figure 15** Corresponding shear stress on the rigid boundary with respect to non-dimensional time at different elasticities of contrast agent when $h^* = 4.0$ using EEM model for the encapsulation

To study the effect of standoff distance on the contrast agent behaviour, figure 16(a) shows the non-dimensional volume of the contrast agent with respect to time at different standoff distances using EEM. As it is observed the bubble shows more expansion when it is initially located further far from the wall since at far distances the contrast agent has more freedom to expand. Figure 16(b) shows the non-dimensional volume centroid of the contrast agent with respect to non-dimensional time using EEM for the encapsulation. At lower standoff distances, the cell membrane has more influence on the bubble which results in the increase of secondary Bjerkness force causing the contrast agent have more translation toward the cell membrane.

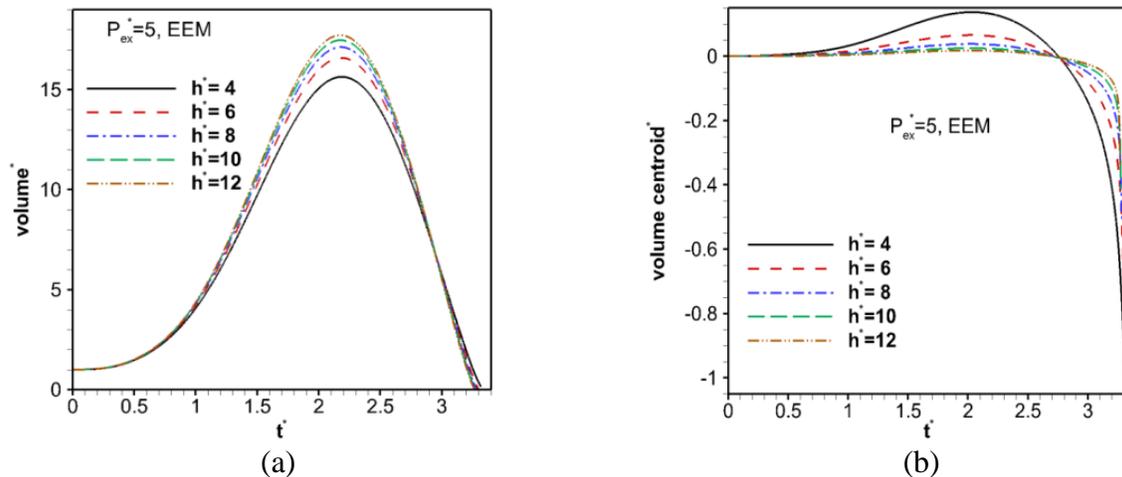

(a)            (b)

**Figure 16** (a) Non-dimensional volume of the contrast agent, (b) non-dimensional volume centroid of contrast agent when $P_{ex}^* = 5.0$ using EEM model for the encapsulation

Figures 17(a) shows the last stage of the collapse phase of the contrast agent located at different initial distances from the wall using both EEM and MM for the encapsulation. As it is seen the bubble shrinks to smaller size at higher standoff distances. Also, it is seen that the width of the bubble jet is lower at higher standoff distances.

Figure 17(b) shows the non-dimensional jet tip velocity of the contrast agent at the axis of symmetry with respect to non-dimensional time for both EEM and MM. It is observed that the jet velocity increases with the increase of standoff distance. Wang et al. have also observed the increase of jet velocity with the increase in standoff distance using Church-Hoff model for the encapsulation (Wang, Manmi et al. 2015). Figure 17(c) show the shear stress on the cell membrane using EEM for the encapsulation. As it is shown the shear stress on the cell membrane is increasing with the increase of standoff distance while the period where the wall is experiencing shear stress is decreasing.

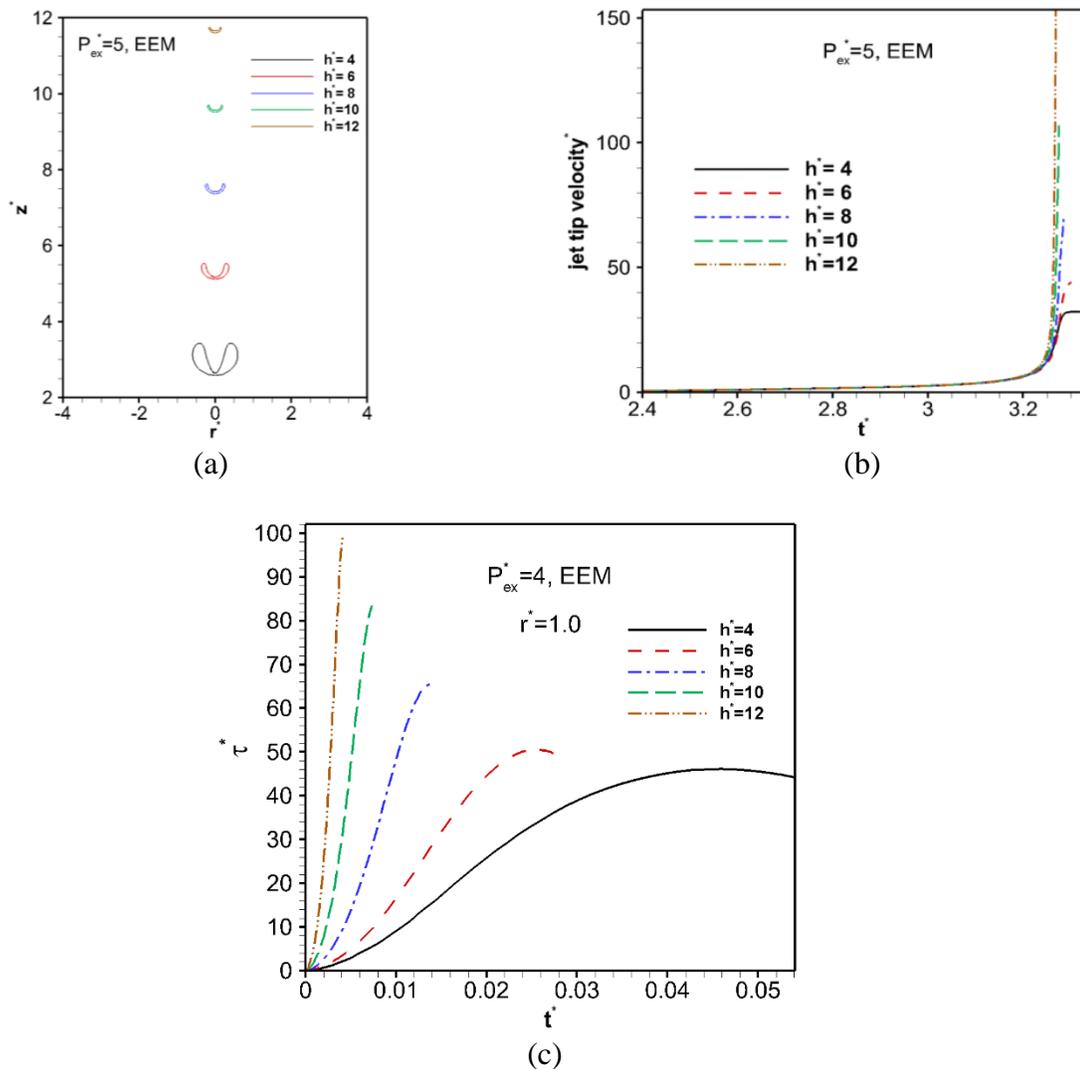

**Figure 17** (a) Last stage of microbubble collapse for different distances of the contrast agent from wall, (b) the velocity of the tip of the jet and (c) the corresponding shear stress on the rigid boundary with respect to non-dimensional time when $P_{ex}^* = 5.0$ using EEM model for the encapsulation

## Conclusion

Micron sized gas-bubbles coated with a stabilizing shell of lipids or proteins, are used as contrast enhancing agents for ultrasound imaging. However, they are increasingly being explored for novel applications in drug delivery through a process called sonoporation, the reversible permeabilization of the cell membrane. Under sufficiently strong acoustic excitations, bubbles form a jet and collapse near a wall. The jetting of free bubbles has been extensively studied by boundary element method (BEM). Here, we implemented interfacial rheological models of the shell into BEM and investigated the jet formation. Two interfacial rheology models have been used to simulate the interfacial rheology of the microbubble encapsulation. The code has been carefully validated against past results. The dynamics of the contrast microbubble and the surrounding fluid is studied varying the shell rheology and other relevant parameters of the problem. Increasing shell elasticity decreases the maximum bubble volume and the collapse time, while the jet velocity increases. The shear stress on the wall is computed and analysed. A phase diagram as functions of excitation pressure and wall separation describes jet formation. Effects of shell elasticity on the phase diagram is investigated.